\documentclass[aps%,twocolumn,
floats,prd,nofootinbib,superscriptaddress,10pt]{revtex4-1}
\usepackage[utf8]{inputenc}
\usepackage{bm}
\usepackage{amsmath}
\usepackage{comment}
\usepackage{color}

\def\ao{\textcolor{black}}
\usepackage{graphicx,amsmath,amsfonts,amssymb%,slashed
}

\begin{document}
\begin{comment}
\begin{center}
    \begin{tabular}{ | l | l | l | l |l|l|l|l|}
    \hline
    $x_0$ & $\xi$ & $\lambda$ & $N_{\mathrm{eff,GW}}$&$x_0$ & $\xi$ & $\lambda$ & $N_{\mathrm{eff,GW}}$ \\ \hline
    0.1 & 1 & 0.01 & 1.50&0.1&1&0.005&1.49 \\ \hline
    0.1 & 3 & 0.01 & 0.0484&0.1&3&0.005&0.0461 \\ \hline
    0.5 & 1 & 0.01 & 1.35&0.5&1&0.005&1.33 \\
    \hline
    0.5&3&0.01&0.0328&0.5&3&0.005&0.0302\\
    \hline
    \end{tabular}
\end{center}
\end{comment}
%\begin{comment}

%\end{comment}
%%%%%%%%%%%%%%%%%%%%%%%%%%%%%%%%%%%%%%%%%%%%%%%%%%%%%%%%%%%%%%%%%%%%%
\title{Limits on primordial black holes from  $\mu$ distortions in cosmic microwave background
 %\\and their dependencies on non-Gaussianity
}
%%%%%%%%%%%%%%%%%%%%%%%%%%%%%%%%%%%%%%%%%%%%%%%%%%%%%%%%%%%%%%%%%%%%%
%\title{Reheating preceded by spinodal instabilities of the Higgs \\and 
%gravitational creation of gravitons}

\author{Tomohiro Nakama}

\affiliation{Department of Physics and Astronomy, Johns Hopkins
     University, 3400 North Charles Street, Baltimore, MD 21218, USA}

\author{Bernard Carr}

\affiliation{School of Physics and Astronomy, Queen Mary University of London,\\
Mile End Road, London, E1 4NS, United Kingdom}

\affiliation{Research Center for the Early Universe (RESCEU), Graduate School of Science, University of Tokyo, Tokyo 113-0033, Japan}

\author{Joseph Silk}

\affiliation{Department of Physics and Astronomy, Johns Hopkins
     University, 3400 North Charles Street, Baltimore, MD 21218, USA}
     
\affiliation{Institut d'Astrophysique de Paris, UMR 7095 CNRS,
Sorbonne Universit\'e, 98 bis Boulevard Arago, 75014 Paris, France
}

\affiliation{BIPAC, Department of Physics, University of Oxford, Keble Road, Oxford OX1 3RH,  United Kingdom}

\begin{abstract}

If primordial black holes (PBHs) form directly from inhomogeneities in the early Universe, then the number in the mass range $10^5 -10^{12}M_{\odot}$ is severely constrained by upper limits to the $\mu$ distortion in the cosmic microwave background (CMB). This is because inhomogeneities  on these scales will be dissipated by Silk damping in the redshift interval $5\times 10^4\lesssim z\lesssim2\times 10^6$.
If the primordial fluctuations on a given mass scale have a Gaussian distribution and PBHs form on the high-$\sigma$ tail, as in the simplest scenarios, then the $\mu$ constraints exclude PBHs in this mass range from  playing any interesting cosmological role. Only if the fluctuations are highly non-Gaussian, or form through some mechanism unrelated to the primordial fluctuations, can this conclusion be obviated. 
%On scales above $10^9M_{\odot}$ the strongest PBH constraint  comes from $y$-distortions in the CMB but there are many other cosmological sources of this.
\end{abstract}

\maketitle

\section{Introduction}

Primordial black holes (PBHs) have been a focus of great interest for nearly 50 years \cite{1967SvA....10..602Z,Hawking:1971ei,Carr:1974nx}, despite there still being no definite evidence for them. One reason for this is that only PBHs could be small enough for Hawking radiation to be important \cite{Hawking:1974rv}, those smaller than about $10^{15}$g  having evaporated by now with many interesting cosmological consequences \cite{Carr:2009jm}. 
Recently, however, attention has shifted to PBHs larger than $10^{15}$g, which  are unaffected by Hawking radiation. 
This is because of the possibility that they provide the dark matter, an idea that goes back to the earliest days of PBH research \cite{1975Natur.253..251C} and has been explored in numerous subsequent works \cite{barrau,Carr:2016drx,Frampton,clesse}.
Since PBHs formed in the radiation-dominated era, they are not subject to the well-known big bang nucleosynthesis (BBNS) constraint that baryons can have at most $5\%$ of the critical density \cite{Cyburt:2003fe}, which is well below the $25\%$ associated with the dark matter. They should therefore be classed as nonbaryonic and, from a dynamical perspective, behave like any other cold dark matter (CDM) candidate. 
There is no compelling evidence that PBHs provide the dark matter, but nor is there evidence for any of the more traditional CDM candidates, 
either from  direct searches  with  underground detectors and particle accelerators or from indirect searches for the expected gamma-ray, neutrino or positron signatures \cite{valentino}. 

Even if nonevaporating PBHs  do not provide all  the dark matter, they could still have 
interesting cosmological effects. For example, they have been invoked to explain 
the heating of the stars in our Galactic disc \cite{1985ApJ...299..633L}, the seeding of the supermassive black holes in galactic nuclei \cite{Bean:2002kx,Kawasaki,carr-silk}, the generation of large-scale structure through Poisson fluctuations \cite{Afshordi:2003zb,carr-silk} and  the associated generation of an infrared background \cite{Kashlinsky:2016sdv},  the reheating and ionization  of the Universe \cite{Ricotti:2007au,Ali-Haimoud:2016mbv}, and the production of  r-process elements \cite{fuller}. More recently it has been proposed that coalescing PBHs could explain the LIGO gravitational wave bursts \cite{Bird:2016dcv}, although this may only require a small fraction of the dark matter to be in PBHs \cite{Sasaki:2016jop}. 
The detection of  four black holes with mass around 30$\rm M_\odot$  has come as a surprise to stellar evolution modelers, so it is natural to consider more exotic types of black holes. 
The suggestion that LIGO could detect gravitational waves from a population of binary intermediate-mass black holes  was originally proposed in the context of the Population III scenario by Bond and Carr \cite{1984MNRAS.207..585B}, and - rather remarkably - a paper in 2014 predicted a Population III coalescence peak at $30M_{\odot}$ \cite{kinugawa}. 
%This is now regarded as unlikely, 
Since Population III stars are baryonic, 
%and therefore subject to the BBNS constraint, 
such black holes could not provide the dark matter, but this would not preclude intermediate-mass PBHs from doing so.
There have been a large number of recent papers on this topic, but the suggestion that there could be a stochastic background of gravitational waves from PBHs goes back a long way \cite{1980A&A....89....6C,Nakamura}.

There are other possible explanations for these effects, 
so they do not necessarily require the existence of PBHs. 
However, all these effects can be used to place interesting constraints on the number of PBHs, and this in turn  constrains the cosmological models which generate them. 
The constraints are most usefully expressed as limits on the fraction $f(m)$ of the dark matter in PBHs of mass $m$
and have recently been summarized in Refs.~\cite{Carr:2009jm} and \cite{Carr:2016drx}. Taken together, they suggest 
%The most stringent ones come from lensing,  dynamical and accretion effects and suggest 
that there are only a few mass windows where PBHs could provide all the dark matter ($f = 1$):
%The most interesting is
 the intermediate-mass range ($10-100\,M_{\odot}$), 
 the lunar-mass range ($10^{20}-10^{24}$g),
 the asteroid-mass range ($10^{16}-10^{17}$g) and Planck mass relics of evaporation ($10^{-5}$g).
% but these are not relevant to the present considerations. 
Even some of these windows may be excluded for a monochromatic PBH mass function, so recently it has been suggested that the PBH dark matter proposal may require an extended mass function \cite{Carr:2016drx}. However, there is some dispute over whether this helps or hinders the proposal for the various mass windows \cite{Green:2016xgy,Kuhnel:2017pwq,Carr:2017jsz}.  
%Although each of the limits comes with various caveats, 
In any case, it seems clear that all the dark matter could  comprise PBHs only if they were smaller than about $10^2 M_{\odot}$. 

%However, this does not preclude PBHs from comprising 
PBHs larger than this might still have an appreciable cosmological density,
%may still comprise a small fraction of the dark matter,
% in higher mass ranges, 
so it is important to consider whether there is a maximum possible mass for a PBH. Since a PBH forming at a time $t$ after the big bang is expected to have  a mass of order the particle horizon size $\sim 10^5(t/s)M_{\odot}$, this depends on how late a PBH can form \cite{Carr:1975qj}.  It is sometimes argued that this
should be before weak freeze-out at $1$~s, corresponding to a maximum mass of  $10^5M_{\odot}$, since otherwise BBNS would be affected. This is because PBH production usually requires large inhomogeneities and this might be expected to disturb the usual BBNS scenario \cite{Cyburt:2008up}.
 However, this argument is not clear cut  because the fraction of the Universe in PBHs at a time $t$ after the big bang  is only $\sim 10^{-6} \Omega_{\mathrm{PBH}} (t/s)^{1/2}$, where $\Omega_{\mathrm{PBH}}$ is the current PBH density in units of the critical density \cite{Carr:1975qj}, so this would be at most $10^{-6}$ at weak freeze-out. 
 
Various limits can be imposed on the density of such large PBHs. As reviewed in Ref.~\cite{Carr:2016drx}, the microlensing of stars has been sought in a wide variety of contexts, and this limits $f(M)$ over various mass ranges below about $10 M_{\odot}$. Above this mass, numerous dynamical effects come into play. In particular, PBHs in the intermediate-mass range would disrupt wide binaries in the Galactic disc. It was originally claimed that this would exclude objects above $400\,M_{\odot}$ \cite{Quinn:2009zg}. However, more recent studies may reduce this mass \cite{Monroy-Rodriguez:2014ula}, so the narrow window between the microlensing and wide-binary bounds is shrinking. Nevertheless, this suggestion is topical because PBHs in the IMBH range  might also explain the sort of massive black hole mergers observed by LIGO.
%There have been a large number of recent papers on this topic but the suggestion that there could be a stochastic background of gravitational waves from PBHs goes back a long way \cite{1980A&A....89....6C,Nakamura}.
More recent studies have examined the induced gravitational wave background  from the fluctuations with amplitude below the PBH threshold \citep{Nakama:2016gzw, garciaPU}.
%, such a backgtound being constrained  by Pulsar Timing Array observations.

Could such huge PBHs be {\it expected} to form? 
%As reviewed in Ref.~\cite{Carr:2005zd}, 
Only a very low production efficiency is needed to generate a significant PBH 
%contribution to the DM 
density today, and they may be generated by three mechanisms~\cite{Carr:2005zd}: (i) the collapse of large-amplitude inhomogeneities \cite{Carr:1975qj}; (ii) a temporary softening of the equation of state (even if the inhomogeneities are small) 
%{\color{red} it having long been realised that an early matter-dominated phase would inevitably produce PBH
\cite{Khlopov:1980mg,Jedamzik:2010dq,Harada:2016mhb,ctv}; (iii) some form of cosmological  phase transition~\cite{Carr:2009jm}. The last two mechanisms are unlikely to be relevant after $1$~s, but the first mechanism could be. \ao{For example, the inflationary models discussed in Refs. \cite{Ivanov:1994pa,Yokoyama:1995ex,Bullock:1996at,Yokoyama:1998pt,hybrid, kawasaki} could produce a spike in the power spectrum of density fluctuations at a mass scale which is not uniquely specified.}
%can be changed.is essentially arbitrary.} 
It could be tuned to the  intermediate-mass range ($10-10^3 M_{\odot}$) if one wants to explain the dark matter \cite{Frampton:2010swl}, but it could, in principle, be much larger.
Another interesting formation mechanism, involving the collapse of inflationary bubbles, has recently been explored in Refs. \cite{Garriga:2015fdk,vilenkin}. 
% in order to explain the dark matter \cite{frametal}.

What would be the possible cosmological consequences of a relatively small density of large PBHs? It has been claimed that some observational anomalies may require the existence nonlinear structures early in the history of the Universe \cite{Dolgov:2016qsm}. For example, it is now known that most galactic nuclei  contain supermassive black holes (SMBHs), extending from around $10^{6}M_\odot$ to $10^{10}M_\odot$ and already in place by a redshift of about $10$ \cite{kormendy}. However, it is hard to understand how such enormous black holes could have formed so early unless there were already large seed black holes well before galaxy formation, with these subsequently growing through accretion.
% \cite{Bean:2002kx}.
 Indeed, pregalactic SMBHs might help to {\it explain} galaxy formation, either by acting as condensation nuclei on account of their gravitational Coulomb effect \cite{hoyle} or through the Poisson fluctuations in their number density \cite{Meszaros:1975ef}.  
As discussed by Carr and Silk \cite{carr-silk}, these seeds might well be PBHs.

Another interesting consequence -- and the main focus of this paper --
arises because the formation of PBHs through mechanism (i) requires density fluctuations, and the dissipation of these fluctuations after $10^6$s by Silk damping generates a $\mu$ distortion in the CMB spectrum \cite{Silk:1967kq}. This applies to fluctuations which fall within the photon diffusion scale 
during the redshift interval $5\times 10^4<z<2\times 10^6$, and gives an upper limit $\delta (M) < \sqrt{\mu} \sim 10^{-2}$ over the mass range $10^4 < M/M_{\odot} < 10^{13}$. 
When the PBH formation probability is relatively large, the dispersion of primordial fluctuations is also expected to be large, so, in principle, the observational limits on the $\mu$ distortions translate into upper limits on the PBH abundance.
This constraint was first mentioned in Ref.~\cite{Carr:1993aq}, based on a result in Ref.~\cite{barrow}, but the limit on $\mu$  is now much stronger \cite{
%Chluba:2012we,
Chluba:2015bqa}. 

It should be stressed that this is a limit on the fluctuations from which the PBHs derive, and it can only be translated into a limit on the PBHs themselves if one assumes a model for their formation.
If the fluctuations are Gaussian and the PBHs form on the  high-$\sigma$ tail, as in the simplest scenario \cite{Carr:1975qj}, one can infer a constraint on $f(M)$ 
%in the range $10^3 < M/M_{\odot} < 10^{12}$.
over a wide mass range. Indeed, a  few years ago Kohri {\it et al.} \cite{Kohri:2014lza} obtained  an upper limit of $10^5 M_{\odot}$ based on this  assumption.
However, the Gaussian assumption may be incorrect. For example, Nakama {\it et al.} \cite{Nakama:2016kfq} 
have proposed a ``patch'' model, in which the relationship between the background inhomogeneities 
and the overdensity in the tiny fraction of the volume which collapses to PBHs is modified.  Also, the production mechanism advocated in Refs.~\cite{Garriga:2015fdk,vilenkin} may entail no density perturbations outside the PBHs at all. The $\mu$-distortion constraint could thus be much weaker,  
depending on the degree of non-Gaussianity of the primordial fluctuations. 
A phenomenological description of such non-Gaussianity was introduced in Ref.~\cite{Nakama:2016kfq} and involves 
a parameter $p$, such that -- for a fixed PBH formation probability -- the dispersion of the primordial fluctuations becomes smaller  as $p$ is reduced from its Gaussian value of $2$, thereby reducing the $\mu$ distortion. 

None of these previous works used the $\mu$ limits to directly constrain the PBH mass fraction, so it is not clear whether this  precludes the low PBH density  required in 
%the seed scenario
some cosmological proposals. \footnote{The mass range associated with the $\mu$ limit is indicated in Fig. 1 of Ref. \cite{Inomata:2016rbd} but not the limit on $f(M)$ itself.}  
In this paper we address this problem by calculating the constraints on $f(M)$ explicitly, using both the FIRAS limit of $\mu=9 \times 10^{-5}$ \ao{\cite{Fixsen:1996nj}} and  the projected upper limit of $\mu<3.6\times 10^{-7}$ from PIXIE~\cite{Abitbol:2017vwa}.
We find that
the $\mu$ distortion is predominantly determined by fluctuations with 30~Mpc$^{-1}<k<~$5000~Mpc$^{-1}$, and this corresponds to the PBH mass range of $10^5M_\odot<M<10^{10}M_\odot$.
The mass range around $10^9M_\odot$ is especially restricted 
and we would need huge non-Gaussianity if such massive PBHs 
were to evade the $\mu$-distortion constraints.
% It would therefore  be more plausible to invoke 
Alternatively, one could argue that the PBHs formed with initial mass below $10^4M_\odot$ and then underwent substantial accretion. The $\mu$ constraint would then be avoided altogether.
%, although there would be another type of $\mu$ {\color{blue}} constraint associated with the accretion \cite{Ricotti:2007au}.

The plan of this paper is as follows: In Sec.~II we derive the form of the $\mu$ constraint on the fraction of the dark matter in PBHs on the assumption  that they form from  Gaussian or non-Gaussian primordial fluctuations with a monochromatic power spectrum. In Sec.~III we extend the analysis to include the $y$-distortion limit, and we discuss how our conclusions depend on the type of non-Gaussianity. We also discuss briefly how our conclusions are modified if the fluctuations are nonmonochromatic or the PBHs accrete, both of which would be expected in a more realistic scenario. We summarize our conclusions in Sec.~IV.
\newpage
\ao{
\section{Dependence of PBH limits from $\mu$ on non-Gaussianity}
}

Primordial fluctuations over a wide range of mass scales will be dissipated by Silk damping when they fall within the photon diffusion scale. 
The diffusion scale at any epoch is the geometric mean of the horizon mass and the mass of unit optical depth to Thompton scattering:
\begin{equation}
M_{\rm D} \sim \sqrt{M_{\tau}M_{\rm H}} \sim 
\begin{cases}
10^{10}(t/t_{eq})^{7/4} M_{\odot}
& (t < t_{eq} ) \\        
10^{13}(t/t_{dec})^{11/6} M_{\odot}
& (t_{eq} < t < t_{dec} ) \, .
\end{cases}
%10^{13}(t/t_{eq})^{7/4} \quad (t < t_{eq}) \, .
\end{equation} 
%this being the geometric mean of the horizon mass and the mass of unit optical depth to Thompton scattering. 
The masses here 
%refer to the matter content but in the present context they 
refer to the radiation content, which is much larger than the matter content before the time of matter-radiation equality ($t_{eq} \sim 10^{10}$s) . 
%However since a region collapsing to a PBH is radiation-dominated. 
The value of $10^{13} M_{\odot}$ at decoupling ($t_{dec} \sim 10^{12}$s) corresponds roughly to the Silk mass, the matter and radiation densities then being comparable.

The associated heat production could affect the CMB in various ways, depending on the epoch, so we first discuss these more general limits.
Photons generated by the dissipation of fluctuations before $t_1=7 \times 10^6$s 
%(the freeze-out time for double-Compton scattering) 
will be completely thermalized, leading to an increase in the photon-to-baryon ratio $S$ of the CMB {through acoustic reheating \cite{Nakama:2014vla}}. In principle, this places constraints on fluctuations on scales below $10^5M_{\odot}$  since $S$ cannot increase much after BBNS, but these limits are relatively weak. Photons generated by the dissipation of fluctuations between $t_1$  and $t_2 \approx 3 \times 10^9$s 
%(the freeze-out time for single-Compton scattering?) 
will lead to a $\mu$ distortion 
%(non-zero chemical potential) 
in the CMB. Photons generated by the dissipation of fluctuations between $t_2$ and $t_3 \approx 3 \times 10^{12}$s (decoupling) will lead to a $y$ distortion. 
%The $\mu$ and $y$ distortions therefore constrain PBHs in the mass ranges $10^5 - 10^9 M_{\odot}$ and $10^9 - 10^{13} M_{\odot}$, respectively.  
%The scale of  $10^{13} M_{\odot}$  corresponds roughly to the Silk mass (the diffusion scale at decoupling) since the matter and radiation densities are comparable then. 
Note that these effects can also be used to limit the number of evaporating PBHs~\cite{tashiro}, those in  the mass ranges $10^{11} - 10^{12}$g and $10^{12} - 10^{13}$g evaporating in the periods $t_1 < t < t_2$ 
and $t_2 < t < t_3$, respectively. The source of the photons is different -- Hawking radiation versus dissipation of fluctuations -- but the interaction with the CMB is the same. 

Fluctuations on a given comoving scale produce PBHs when that scale enters the horizon, and only much later are  fluctuations on the same scale dissipated. To find the PBH mass ranges relevant to the three types of distortion mentioned above, we must find the PBH mass $M$ corresponding to the diffusion mass $M_{\rm D}$. These masses are different because a PBH forms when the Universe is radiation dominated. Thereafter its mass remains constant but the radiation mass of a comoving region outside the black hole decreases as $a^{-1}$, which scales as $ t^{-1/2}$ for $t < t_{eq}$ and  $ t^{-2/3}$ for $t > t_{eq}$. One can show that the PBH mass associated with the diffusion mass $M_{\rm D}$ is
\begin{equation}
M \sim 
\begin{cases}
10^{2} (M_{\rm D}/M_{\odot})^{6/7} M_{\odot}
& (t < t_{eq} ) \\        
10 (M_{\rm D}/M_{\odot})^{10/11} M_{\odot}
& (t_{eq} < t < t_{dec} ) \, .
\end{cases}
\end{equation}
This shows that  the PBH mass ranges constrained by the $S$, $\mu$ and $y$ observations are $M/M_{\odot} < 10^5$, $10^5 < M/M_{\odot} < 10^{11}$ and $10^{11} < M/M_{\odot} < 10^{13}$, respectively.
%, so this is also the mass scale on which the PBHs are constrained.
%, {providing the masses in the expression for  $M_{\rm D}$ are interpreted as  referring to the radiation content}.
% \ao{at horizon reentry}.

We now focus on the $\mu$ distortions.
In Ref. \cite{Nakama:2016kfq}, 
the following phenomenological description of the non-Gaussian probability density function (PDF) of the curvature perturbation $\zeta$ was introduced:
\begin{equation}
P(\zeta)=\frac{1}{2\sqrt{2}\tilde{\sigma} \Gamma\left(1+1/p\right)}\exp \left[-\left(\frac{|\zeta |}{\sqrt{2}\tilde{\sigma}}\right)^p\right].\label{pdf2}
\end{equation}
This satisfies $\int_{-\infty}^\infty P(\zeta)d\zeta =1$ and reduces to  the Gaussian distribution of Ref.~\cite{Carr:1975qj} when $p=2$. 
The dispersion is 
\begin{equation}
\sigma^2\equiv \int_{-\infty}^\infty \zeta^2 P(\zeta)d\zeta=\frac{2\Gamma(1+3/p)}{3\Gamma(1+1/p)}\tilde{\sigma}^2,
\end{equation}
where $\Gamma(a)$ is the gamma function. 
In particular, $\sigma=\tilde{\sigma}$ when $p=2$, as expected.

 %[CITE OTHER LITERATURE ON PBH FORMATION FROM NON-GAUSSIAN FLUCTUATIONS.] 
We estimate the fraction of the Universe collapsing into PBHs to be
\begin{equation}
\label{beta}
\beta=\int_{\zeta_c}^\infty P(\zeta)d\zeta =\frac{\Gamma(1/p, 2^{-p/2}(\zeta_c/\tilde{\sigma})^p)}
{2p\Gamma(1+1/p)},
\end{equation}
where $\zeta_c$ is the threshold for PBH formation and $\Gamma(a,z)$ is the incomplete gamma function.
 %[PRESUMABLY THIS CAN BE EXPRESSED IN TERMS OF THE ``erfc'' FUNCTION. 
For $p=2$ this reduces to 
\begin{equation}
\beta=2^{-1}\mathrm{erfc}(2^{-1/2} \zeta_c/\sigma) \, .
\label{gauss}
\end{equation}
%Equation~(\ref{beta}) 
This shows that increasing $\mu$ sensitivity is equivalent to increasing $p$.  Our results are very sensitive to the value of $\zeta_c$, but there is some ambiguity about this since it depends upon the perturbation profile, pressure gradients playing an important role \cite{Nakama:2013ica}. Therefore the threshold can only be specified in terms of some range.  If we use the value of $\zeta$ at the peak of the perturbation, $\zeta_c$ lies in the range 0.67-1.05 according to numerical simulations \cite{Harada:2017fjm}. 
But  this peak value is subject to what is called the environmental effect \cite{harada4,young}. Harada \textit{et al}. \cite{Harada:2017fjm} suggest the range 0.95-1.26 by approximately converting the threshold in terms of the density perturbation in the comoving slice
%, obtained from numerical simulations, 
to $\zeta$.
 %suggest the range 0.95-1.26 but this depends on the conversion between $\zeta$ and the density perturbation in the comoving slice, which is problematic in the non-linear regime. 
To make our limits as conservative as possible, we use the threshold $\zeta_c\simeq 0.67$ obtained for the class of perturbation profiles considered in 
%We use the threshold $\zeta_c\simeq 0.67$ for PBH formation \cite{Harada:2017fjm}. This is the value obtained for a class of perturbation profiles in previous numerical soimulations \cite{Nakama:2013ica,Nakama:2014fra}  See also the analytical model of Ref. \cite{Harada:2013epa} as well as
%previous numerical simulations 
 previous numerical simulations \cite{Harada:2017fjm}.
% for a more general discussion. 
%but see also Refs. \cite{Nakama:2013ica,Nakama:2014fra} for numerical simulations}.
Equation~(\ref{beta}) can be inverted to give
%solved for $\tilde{\sigma}$ as 
\begin{equation}
\tilde{\sigma}=\frac{2^{-1/2}\zeta_c}{[Q^{-1}(1/p,2\beta)]^{1/p}} \, ,
\end{equation}
where $Q^{-1}(a,z)$ is the inverse of the regularized incomplete gamma function $Q(a,z)\equiv \Gamma(a,z)/\Gamma(a)$, 
so that $z=Q^{-1}(a,s)$ if $s=Q(a,z)$. 

Let us consider the following {dimensionless} delta-function
%-type 
power spectrum: 
\begin{equation}
{\cal P}_\zeta=\sigma^2 k\delta(k-k_*),\label{deltafunction}
\end{equation}
which leads to the $\mu$ distortion 
\cite{Chluba:2012we}
\begin{equation}
\mu\simeq 2.2\sigma^2 \left[
\exp\left(-\frac{\hat{k}_*}{5400}\right)
-\exp\left(-\left[\frac{\hat{k}_*}{31.6}\right]^2\right)
\right],\label{mu}
\end{equation}
where $\hat{k}_*$ is the wave number in units of Mpc$^{-1}$.
%$k_*=\hat{k}_*\rm{Mpc}^{-1}$. 
%We adopt $\mu_{\rm{upper}}=9\times 10^{-5}$ as a 2$\sigma$ upper limit obtained by COBE/FIRAS \cite{Fixsen:1996nj}. 
%along with the COBE/FIRAS upper limit. 
The wave number and the  PBH mass  are related via \cite{Nakama:2016gzw}
\begin{equation}
k\simeq 7.5\times 10^5 \gamma^{1/2}\mathrm{Mpc}^{-1}\left(\frac{g}{10.75}\right)^{-1/12}\left(\frac{M}{30M_\odot}\right)^{-1/2},\label{kmass}
\end{equation}
where $\gamma$ gives the size of the PBH in units of
the horizon mass at formation and $g$ is the  number of degrees of freedom of relativistic particles. 
We set $\gamma=1$ and $g=10.75$ for all masses for simplicity.\footnote{Strictly speaking, $g$ should be 
%taken to be 
somewhat smaller for $M>10^5M_\odot$, but this simplification does not cause significant errors due to the weak dependence on $g$ in Eq.~(\ref{kmass}).} The initial abundance $\beta$ is related to $f=\Omega_{\mathrm{PBH}}/\Omega_{\mathrm{DM}}$, where $\Omega_{\mathrm{DM}} \simeq0.27$ \cite{Ade:2015xua}, via \cite{Nakama:2016gzw} 
\begin{equation}
\beta\simeq 1.1\times 10^{-8}\gamma^{-1/2}\left(\frac{g}{10.75}\right)^{1/4}\left(\frac{\Omega_{\mathrm{DM}}}{0.27}\right)^{-1}\left(\frac{M}{30M_\odot}\right)^{1/2}f.
\label{betaf}
\end{equation}
This is just Eq. (2.5) of Ref.~\cite{Carr:2009jm} with $M$ normalized to the mass $30 M_{\odot}$ indicated by the LIGO events. Then, using both the FIRAS limit ($\mu<9\times 10^{-5}$) \cite{Fixsen:1996nj} and the projected upper limit of 
%$\mu<3.6\times 10^{∁E}$ 
$\mu<3.6\times 10^{-7}$
for PIXIE \cite{Abitbol:2017vwa}, the above phenomenological model of non-Gaussianity can be used to calculate the $\mu$-distortion constraints on $\beta$ or $f$. The results for $f$ are presented in Fig.~1; the results  for $\beta$ are not shown explicitly, but they have a similar form and can be inferred from Eq.~(\ref{betaf}). The figure shows that the limits are  highly restrictive but much less so \ao{in the non-Gaussian cases}
 %shown there} 
than the Gaussian one. 
%For comparison, we show a simple extrapolation of the accretion constraint of Ref.~\cite{Poulin:2017bwe}.
The PIXIE constraint may no longer be relevant since the project has not been approved. Therefore 
we give constraints for both PIXIE and a hypothetical future experiment that we dub HYPERPIXIE, assumed to \ao{give $\mu<10^{-9}$.}
%have 100 times the sensitivity of PIXIE.

The term in square brackets in Eq.~(\ref{mu}) peaks with a value of $1$ at $\hat{k}_* = 80$, which corresponds to a mass $2.6\times 10^9M_\odot$, so the $\mu$ distortion is most sensitive to modes on this scale and has a value
%around $80  \,\mathrm{Mpc}^{-1}$, at which 
$\mu \simeq 2.2\sigma^2$. For $\zeta_c=0.67$, this implies $\zeta_c/\sigma\simeq (100,1.7\times 10^3,3.1\times 10^4)$ for $\mu=(9\times 10^{-5},3.6\times 10^{-7},10^{-9})$. Hence
%, at this wavenumber,
the decimal
%common 
logarithm of the PBH formation probability in the Gaussian case,
%for Gaussianity, 
given by Eq.~(\ref{gauss}), 
%$\beta=2^{-1}\mathrm{erfc}(2^{-1/2}\times \zeta_c/\sigma)$, 
formally reaches $(-2.4\times 10^3,-6.0\times 10^5,-2.1\times 10^8)$.
%, which explains the tight limits.
Although this  limit is very strong, there is also a value of $f(M)$ which corresponds to having just one PBH per current Hubble horizon. This has been described as the ``incredulity limit'' \cite{carsak} and can be written as
%  It would be instructive to introduce the fraction $f_1$ of one single PBH of mass $M$ in the current Hubble radius: 
\begin{equation}
f_1=\frac{M}{\Omega_{\mathrm{DM}}M_H}=8.2\times 10^{-14}\left(\frac{\Omega_{\mathrm{DM}}}{0.27}\right)^{-1}\left(\frac{h}{0.67}\right)\left(\frac{M}{10^9M_\odot}\right) \, ,
\end{equation}
where $M_H$ is the current horizon mass,
\begin{equation}
M_H=
%\frac{4\pi}{3}H^{-3}\rho_{\mathrm{cr}}=
\frac{c^3}{2GH}=4.5\times 10^{22}M_\odot\left(\frac{h}{0.67}\right)^{-1}.
\end{equation}
%Our upper limits are very stringent, but note that the values of $f$ below $f_1$ may not make much sense. 
The $\mu$ limit can be well below this, so we also plot $f_1$ in the figures for comparison. 

Figures 2 and 3 show the corresponding results
for the quadratic  and cubic non-Gaussianity investigated in the context of PBHs in Refs. \cite{Byrnes:2012yx,Nakama:2016gzw}:
\begin{equation}
\zeta=\zeta_G+\frac{3}{5}f_{\mathrm{NL}}(\zeta_G^2-\sigma_G^2),
\end{equation}
\begin{equation}
\zeta=\zeta_G+g\zeta_G^3, \quad g\equiv \frac{9}{25}g_{\mathrm{NL}}.
\end{equation}
%[COULD COMBINE INTO A SINGLE EQUATION] 
The dispersion $\sigma^2=\langle\zeta^2\rangle$ is then related to $\sigma_G^2\equiv \langle\zeta_G^2\rangle$ via
\begin{equation}
\sigma^2=\sigma_G^2+2\left(\frac{3}{5}f_{\mathrm{NL}}\right)^2\sigma_G^4,
\end{equation}
\begin{equation}
\sigma^2=\sigma_G^2+6g\sigma_G^4+15g^2\sigma_G^6.
\end{equation}
For each $f_{\mathrm{NL}}$ or $g_{\mathrm{NL}}$ and mass $M$, which can be translated into $k_*$, the dispersion $\sigma^2$ (or equivalently $\sigma_G^2$) corresponding to the upper limit on $\mu$ is obtained. The corresponding %abundance $\beta$ or 
value of $f$ can then be obtained using Refs~\cite{Byrnes:2012yx,Nakama:2016gzw}, which leads to Figs.~2 and 3. 
The limits for cubic non-Gaussianity are
%can be 
weaker than those for quadratic non-Gaussianity, and the limits for  $p-$type non-Gaussianity are
%can be 
even weaker for sufficiently small 
%values of 
$p$. 
%These differences can be understood from Fig.~4, where the values of $\sigma$ as functions of the non-Gaussian parameters are compared, fixing $f=10^{-4}$, which corresponds to $\beta\simeq2\times 10^{-10}$ for $10^6M_\odot$. 
As noted in Ref. \cite{Nakama:2016gzw}, in the limit $f_{\mathrm{NL}}=\infty$ ,we have
\begin{equation}
\beta=\mathrm{erfc[(\tilde{\zeta}_{\mathrm{c}}/2)^{1/2}]}  \quad \rm{with} \quad \tilde{\zeta}_{\mathrm{c}}=\sqrt{2}\zeta_{\mathrm{c}}/\sigma+1
\end{equation}
where $\mathrm{erfc}(x)$ denotes the complementary error function.
 %and $\tilde{\zeta}_{\mathrm{th}}=\sqrt{2}\zeta_{\mathrm{th}}/\sigma+1$. 
In the limit $g_{\mathrm{NL}}=\infty$, we have
\begin{equation}
\beta=\frac{1}{2}\mathrm{erfc}\left[\frac{1}{\sqrt{2}}\left(\frac{\zeta_{\mathrm{c}}}{\sigma/\sqrt{15}}\right)^{1/3}\right].
\end{equation}
These relations and Eqs.~(\ref{mu}) and (\ref{kmass}) lead to the curves of Figs. 2 and 3 with $f_{\mathrm{NL}}=\infty$ and $g_{\mathrm{NL}}=\infty$, which serves as a check. 

%%%%%%%%%%%%%%%%%%%
\begin{figure}[htp]
\begin{center}
\includegraphics[width=8cm,keepaspectratio,clip]{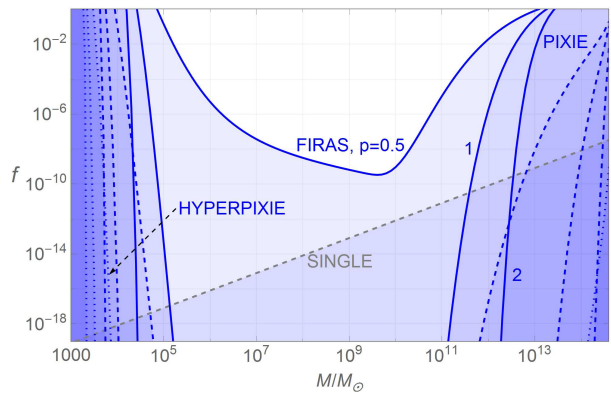}
\end{center}
\caption{
Upper limits on $f=\rho_{\mathrm{PBH}}/\rho_{\mathrm{DM}}$
%$\beta=\rho_{\mathrm{PBH}}/\rho_{\mathrm{rad}}$ at PBH formation 
for different values of the  non-Gaussianity parameter $p$ [Eq. (\ref{pdf2})]. The solid curves are for FIRAS ($\mu=9\times 10^{-5}$) and  correspond to $p=0.5,1,2$ (from top to bottom). The dashed and dotted curves are 
for PIXIE ($\mu=3.6\times 10^{-7}$) and HYPERPIXIE ($\mu=10^{-9}$), respectively, with the same values of $p$ from top to bottom. The SINGLE line corresponds to having one PBH per current Hubble volume.}
%The gray dashed line represents a simple extrapolation of the PBH accretion limit 
%from accretion onto PBHs provided in 
%of Ref. \cite{Poulin:2017bwe}. 
%(Right) Translation into the current fraction of the dark matter in PBHs: $\Omega_{\mathrm{PBH}}/\Omega_{\mathrm{DM}}$.
\label{betap}
\end{figure}
%%%%%%%%%%%%
%%%%%%%%%%%%%%%%%%%

%%%%%%%%%%%%
%%%%%%%%%%%%%%%%%%%
\begin{figure}[htp]
\begin{center}
\includegraphics[width=8cm,keepaspectratio,clip]{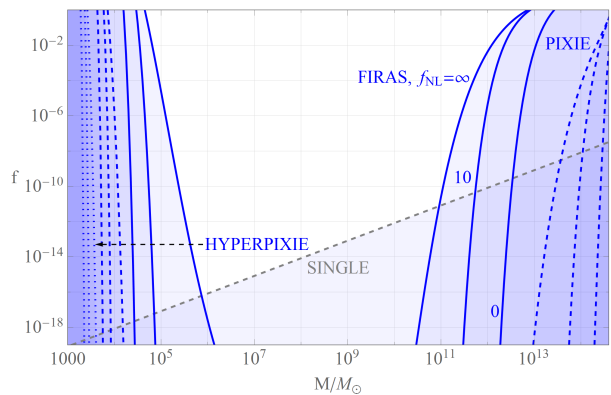}
\end{center}
\caption{
 Upper limits on $f=\rho_{\mathrm{PBH}}/\rho_{\mathrm{DM}}$
%the PBH fraction $\beta=\rho_{\mathrm{PBH}}/\rho_{\mathrm{rad}}$ 
%at formation 
for quadratic non-Gaussianity with $f_{\mathrm{NL}}=\infty,10,0$ from top to bottom. The FIRAS, PIXIE,  HYPERPIXIE and SINGLE curves are as for Fig.~1.}
% The solid curves are for FIRAS ($\mu=9\times 10^{-5}$), the dashed curves 
%for PIXIE ($\mu=3.6\times 10^{-7}$) and the dotted curves for HYPERPIXIE ($\mu=10^{-9}$) with \textcolor{red}{$f_{\mathrm{NL}}=\infty,10,0$} from top to bottom.
%the same values of $f_{\mathrm{NL}}$, . 
%The gray dashed line represents a simple extrapolation of the PBH accretion limit 
%a limit from accretion onto PBHs provided in 
%of Ref. \cite{Poulin:2017bwe}. 
%(Right) Upper limits on $f=\rho_{\mathrm{PBH}}/\rho_{\mathrm{DM}}$ at the present timefor the case of quadratic non-Gaussianity. 
\label{betafnl}
\end{figure}
%%%%%%%%%%%%
%%%%%%%%%%%%%%%%%%%
%%%%%%%%%%%%
%%%%%%%%%%%%%%%%%%%
\begin{figure}[htp]
\begin{center}
\includegraphics[width=8cm,keepaspectratio,clip]{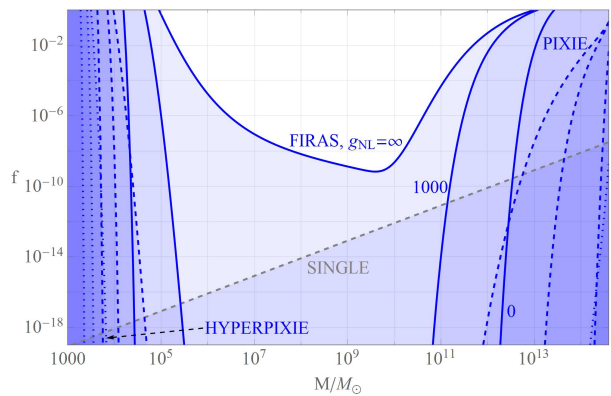}
\end{center}
\caption{
 Upper limits on $f=\rho_{\mathrm{PBH}}/\rho_{\mathrm{DM}}$
%$\beta=\rho_{\mathrm{rad}}/\rho_{\mathrm{PBH}}$ 
for cubic non-Gaussianity with $g_{\mathrm{NL}}=\infty, 1000, 0$ from top to bottom and the same values of $g_{\mathrm{NL}}$. The FIRAS, PIXIE, HYPERPIXIE and SINGLE curves are as for Fig.~1.}
%The solid lines are for FIRAS and from top to bottom \textcolor{red}{$g_{\mathrm{NL}}=\infty, 1000, 0$}.  The dashed and dotted curves are for PIXIE and HYPERPIXIE, respectively, with the same values of $g_{\mathrm{NL}}$.
%(Right) Upper limits on $f=\rho_{\mathrm{PBH}}/\rho_{\mathrm{DM}}$ for the case of cubic non-Gaussianity.  
\label{betagnl}
\end{figure}
%%%%%%%%%%%%

\section{Discussion}

\begin{figure}[htp]
\begin{center}
\includegraphics[width=9cm,keepaspectratio,clip]{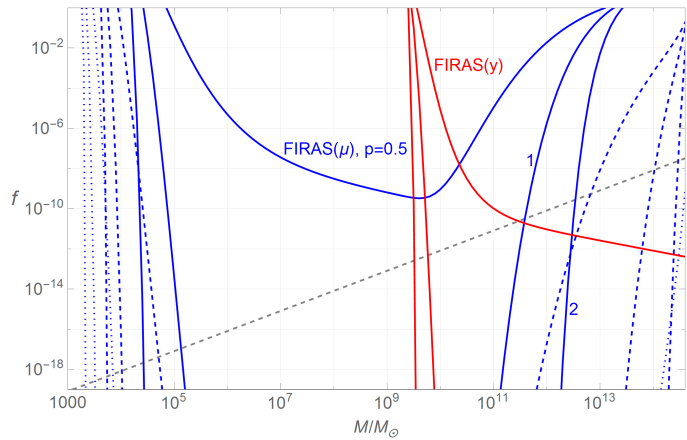}
\end{center}
\caption{
Upper limits on $f=\rho_{\mathrm{PBH}}/\rho_{\mathrm{DM}}$
%$\beta=\rho_{\mathrm{rad}}/\rho_{\mathrm{PBH}}$ 
from $\mu$ (blue) and $y$ (red) FIRAS observations for different values of $p$. The Gaussian case corresponds to $p=2$. Also shown are the PIXIE and HYPERPIXIE $\mu$ limits and the line corresponding to  one PBH per current Hubble volume.
%\ao{(Bottom) Equivalent upper limits on $f=\rho_{\mathrm{PBH}}/\rho_{\mathrm{DM}}$.}
%for the case of the cubic non-Gaussianity.
}  
\label{y}
\end{figure}

Although we have focused mainly on the $\mu$ distortions associated with PBH formation, we should also comment briefly on the $y$ distortions, which can be calculated  using Eq. (9b) in Ref. \cite{Chluba:2012we} and assuming $y<1.5\times 10^{-5}$ \cite{Fixsen:1996nj}.
It is clear from Fig.~\ref{y}, which shows the $\mu$ and $y$ limits \ao{for FIRAS} with different values of $p$, that there is a  transition at around $10^9 M_{\odot}$ above which the $y$ constraint dominates. This scale is independent of the value of $p$ and can be understood from the qualitative discussion at the start of Sec.~II. 
The $y$ curves are interesting, even if one does not expect PBHs larger than $10^9 M_{\odot}$ in practice. \ao{Note that including 
%Although there are many 
other sources of $y$ distortion would merely strengthen the PBH limits. Indeed,} HYPERPIXIE will be looking for the $y$ signal from astrophysical sources at the level of $10^{-6}$ to $10^{-8}$, and such a signal will be obscured if the $y$ from Silk damping exceeds this.
%The \ao{ $\mu$ constraints on PBHs} are shown for both FIRAS (actual) and \ao{HYPERPIXIE (potential) but 
%The $y$ constraints are only shown for FIRAS
%potential future limits} but the $y$ constraint is only shown for FIRAS.  
The total constraint for given $p$ comes from the combination of the $\mu$ and $y$ limits. The value of $f$ at the intersect gives a local maximum in the constraint, but there are extended  minima on either side of this, with the $\mu$ limit being somewhat weaker than the $y$ limit.
 %[WHY?]. 
The value of $f$ at the maximum  decreases as $p$ increases. % and  reaches $10^{-20}$ \textcolor{blue}{[UPDATE?]} at the Gaussian limit of $2$. 
For $p>2$, the constraints would be even tighter than in the Gaussian case. %\textcolor{red}{The value of $f$ for the extended minima also decreases as $p$ increases and in the Gassian limit reaches $10^{-4800}$ for $\mu$ and $10^{-5200}$ for $y$. [UPDATE]} 
%This shows that the simplest (Gaussian) PBH formation scenario is remarkably constrained. %For HYPERPIXIE, the corresponding value of $\sigma$ is about $2\times10^{-5}$, \textcolor{red}{so peaks collapsing to PBHs have $\zeta_c/\sigma\simeq 4.5\times10^4$. The PBH formation probability for Gaussian fluctuations is therefore $2^{-1}\mathrm{erfc}(2^{-1/2}\times \zeta_c/\sigma)\sim10^{-4.3\times10^8}$. [UPDATE]} Although not shown explicitly in the figures, the $\beta$ curves for the Gaussian HYPERPIXIE case have a minimum at roughly this value. 

At the present epoch we require SMBHs with mass $10^8 - 10^9 M_{\odot}$ in galactic nuclei, and the Magorrian scaling  implies $f_{\rm SMBH} \sim 10^{-4}$ \cite{magorrian}. 
%\textcolor{red}{[CITE MAGORRIAN?]} 
If there were no accretion, the PBH scenario would be  excluded by the above analysis unless $p$ were as low as $0.5$ or $f_{\rm NL}$ and $g_{\rm NL}$ as high as 5000. However, 
%the abve analysis has neglected the accretion of the PBHs and this could be appreciable, so we now consider the implications of this.
 the values of $M$ and $f$ indicated in Figs. 1-4 correspond to the {\it formation} epoch of the PBHs, 
%but both $M$ and $f$ could subsequently increase appreciably as a result of accretion. Indeed, 
and  the production of SMBHs in galactic nuclei probably requires a very large accretion factor. For example, for Eddington-limited growth, the initial seed mass could be a smaller by a factor of $10^5 - 10^6$, corresponding to an initial $f$ of $10^{-9} - 10^{-10}$.  If the final  mass is much larger than the initial mass, one expects $f \propto M$, so the value of ($f$,$M$) in Figs. 1-4  evolves along a straight line.
% connecting the non-accretion and Eddington-accretion cases.  
More generally, 
%It should be stressed that the $\mu$ constraint on $f(m)$ only applies if $m$ is interpreted as the {\it formation} mass of the PBH.  The constraint on the {\it current} mass distribution may be weaker if the PBHs grow a lot through accretion. 
since the $\mu$ limit is only important above $10^5M_{\odot}$, one could circumvent it altogether,
%with sufficient accretion. However, we would need 
provided the mass increases by a factor $(m/10^5M_{\odot})^{-1}$.
% for this to happen, 
%The current mass could be much larger repreesnts the formation mass the the black holes accrete so much that their initial mass is too low 
However, in this case the accretion of  gas onto the PBHs may also contribute to temperature and polarization anisotropies and $y$ distortions of the CMB \cite{Ricotti:2007au,Ali-Haimoud:2016mbv,Poulin:2017bwe}, so accretion may not save the scenario. 
\if
%In general, the values of $M$ and $f$ will increase together.
 %One must therefore be careful in relating Figs 1-4 to particular scenarios in which PBHs influence cosmic structure. 
%However, most models for the formation of SMBHs in galactic nuclei require a very large accretion factor. For example, for Eddington-limited growth, the initial seed mass needs to be $10^3 - 10^4 M_{\odot}$, corresponding to a growth factor of  $10^5 - 10^6$ and an initial $f$ of $10^{-9} - 10^{-10}$. 
%These values do not conflict with our constraints even in the Gaussian case. 
%In considering whether the non-Gaussianity required for intermediate accretion scenarios is reasonable, the following feature may be useful: providing the final  mass is much larger than the initial mass, one expects $f \propto M$, so the value of ($f$,$M$) in Figs. 1-4 must evolve along a straight line connecting the non-accretion and Eddington-accretion cases.  
\fi

The above analysis has considered the types of non-Gaussianity associated with the parameters $p$, $f_{\rm NL}$ and $g_{\rm NL}$, so it may be useful to comment on the connection between these. 
The parameter $p$ 
changes the exponential falloff of the PDF, so that it declines as $\exp(-|\zeta |^p)$ rather than $\exp(-|\zeta |^2)$, as in the Gaussian case. This is slower than the Gaussian decline for $p<2$, so for fixed $\sigma$ this widens the PDF but lowers the central height since area is conserved. For $p<2$ and the same value of  $\sigma$, the modified curve goes above the Gaussian curve  at a value of $\zeta$ less than the critical value $\zeta_c = 0.67$ required for PBH formation, so PBH production will be increased. This explains why the constraints on $f$ become weaker as $p$ decreases below 2.  Of course, {this is a particularly simple type of non-Gaussianity, 
and we do not currently know what kind of physics in the early Universe would lead to it; thus, this is just a convenient phenomenological toy model.
Nevertheless, this  illustrates qualitatively why the deviations from Gaussianity  {that we have investigated} weaken the PBH constraints. 

The parameters $f_{\rm NL}$ and $g_{\rm NL}$ are more traditional measures of non-Gaussianity, whose significance is well understood. From Eq. (14), $f_{\rm NL}$ couples with $\zeta_G^2 - \sigma_G^2$, so that the average $ \langle \zeta \rangle$ is zero, while $g_{\rm NL}$ 
 couples with $\zeta_G$ from Eq. (15).  
In principle, one could combine Eqs. (14) and (15) into a single equation to allow
%see what happens if 
$f_{\rm NL}$ and $g_{\rm NL}$ to vary together. 
%; likewise with eqns (11) and (12).
Generally  a pure $f_{\rm NL}$ can yield either a weaker or  stronger constraint than some mixture of $f_{\rm NL}$ and $g_{\rm NL}$, depending on their precise  values.  
%of fNL and gNL of the latter.} 
The relative effects of $p$, $f_{\rm NL}$ and $g_{\rm NL}$ are shown in Fig.~\ref{sigma}, where the values of $\sigma$ as functions of the non-Gaussian parameters are compared for $f=10^{-4}$, corresponding to $\beta\simeq2\times 10^{-10}$ for $10^6M_\odot$. 
%
%[DISCUSSS NON-GAUSSIANITY EXPECTED FOR PARTICULAR PBH INFLATIONARY SCENARIOS. \textcolor{red}{THIS REFERS TO EARLY WORK OF  IVANOV AND HIDALGO -- NOT RECENT WORK OF BYRNES ET AL -- BUT ANY RESERVATIONS ABOUT THE LATTER CAN BE EXPRESSED.]} \\

%%%%%%%%%%%%%%%%%%%
\begin{figure}[htp]
\begin{center}
\includegraphics[width=9cm,keepaspectratio,clip]{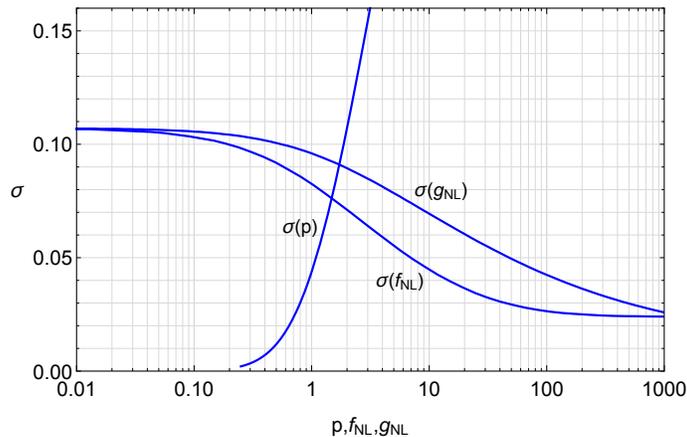}
\end{center}
\caption{
The values of $\sigma$ as a function of the non-Gaussian parameters $p, f_{\mathrm{NL}}$ and $g_{\mathrm{NL}}$ for  $f=10^{-4}$.
%, which corresponds to $\beta\simeq 2\times 10^{-10}$ for $M=10^6M_\odot$. 
The function $\sigma(p)$ can take a wider range of values than $\sigma(f_{\mathrm{NL}})$ and $\sigma(g_{\mathrm{NL}})$. The Gaussian PDF corresponds to $p=2$, $f_{\mathrm{NL}}=0$ and $g_{\mathrm{NL}}=0$. Note that $\sigma(f_{\mathrm{NL}}=\infty)\simeq 0.024$, and the PDF with $p\simeq 0.7$ gives a similar value.
%, but note that 
The PDF is 
not an even function for  quadratic non-Gaussianity, unlike the PDFs specified by $p$ and $g_{\mathrm{NL}}$. Also $\sigma (g_{\mathrm{NL}}=\infty)\simeq 0.01 \simeq \sigma(p=0.5)$, so  the PDFs with $g_{\mathrm{NL}}=\infty$ and $p=0.5$ would look very similar.}
\label{sigma}
\end{figure}
%%%%%%%%%%%%

It should be stressed that there are many other types of non-Gaussianity, so the above analysis should only be regarded as demonstrating the importance of  this property in principle for the form of the $\mu$ constraint on PBHs. In particular, 
%, so  different types and on relationship to more general types of non-Gaussianity.Explain the link with the patch model.} 
Ref. \cite{Nakama:2016kfq} discusses an inflationary scenario -- termed the patch model --  which can produce a sufficient number of PBHs to account for the observed high-redshift quasars, while keeping $\sigma$ sufficiently small to evade constraints from $\mu$ distortions. This model is highly non-Gaussian and essentially introduces a spike in the PDF. This form of non-Gaussianity is very different from the types investigated here, so we have not tried to relate them.

Although we have assumed a delta-function primordial power spectrum,  we can generalize our analysis to the more plausible situation in which the fluctuations have an extended power spectrum.
%we comment on the implications of the PBHs having an extended mass function, since this is more plausible \cite{Carr:2016drx}.}
%We do not discuss them any further here.
%\textcolor{blue}{Though we have used a delta-function-type primordial power spectrum, it would be important to extend our analysis to extended power spectra. 
As a first step, let us consider the following power spectrum:
\begin{equation}
{\cal P}_\zeta=\sigma^2k\frac{1}{\sqrt{2\pi}\sigma_*k_*}\exp\left(-\frac{(k-k_*)^2}{2\sigma_*^2k_*^2}\right).
\end{equation}
This reduces to the delta-function spectrum in the limit $\sigma_*\rightarrow 0.$
The $\mu$ distortion can then 
%for extended power spectra can also
 be calculated using the formalism of Ref. \cite{Chluba:2012we} and, for each $\sigma_*$ and $k_*$, the value of $\sigma$ corresponding to the upper limit on $\mu$ can be obtained. 
%Focusing on 
In the Gaussian case, $\beta$ is calculated using Eq. (\ref{gauss}), and 
%by $\beta=2^{-1}\mathrm{erfc}(2^{-1/2}\times \zeta_c/\sigma)$, which we simply
we translate this to $f$ using the correspondence between $M$ and $k_*$ implied by Eq. (\ref{kmass}).
 %using the mass corresponding to $k_*$. 
The upper limit on $f(M)$ so obtained 
%for each $k_*$, translated to mass,
 is plotted in Fig. \ref{width}. The limits
% on $f$
 for $M\lesssim 10^5M_\odot$ ($M\gtrsim 10^{12}M_\odot$) are tighter when $\sigma_*$ is larger, due to the additional presence of modes with $k<k_*$ ($k>k_*$) which contribute to $\mu$ relatively efficiently but are absent 
in the corresponding case for a delta-function spectrum. 

%We know that such enormous black holes are present in quasars, albeit with a very low cosmological density. Even if they are primordial in origin, the diffuse hard photon background implies they could have experienced significant growth by accretion \cite{soltan}, in which case their initial mass may have been too low for the $\mu$ limit to apply. 

%%%%%%%%%%%%%%%%%%%
\begin{figure}[htp]
\begin{center}
\includegraphics[width=9cm,keepaspectratio,clip]{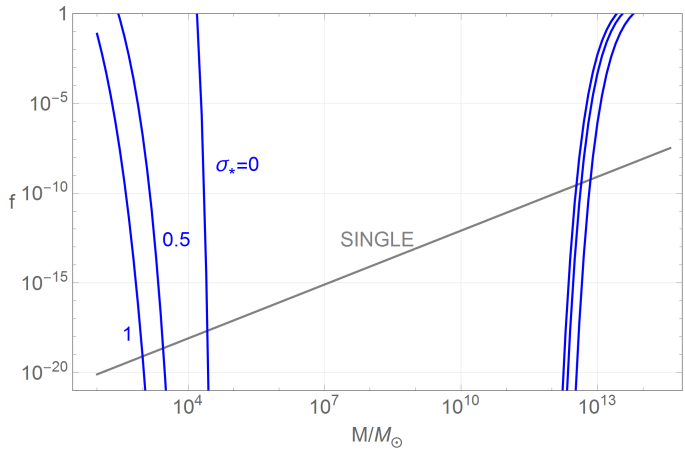}
\end{center}
\caption{
Upper limits on $f$ for extended fluctuation spectra, assuming the FIRAS upper limit $\mu=9\times 10^{-5}$. The width of the spectrum is taken to be $\sigma_*=(0,0.5,1)$. Note that the spectrum is reduced to a delta-function spectrum for $\sigma_*=0$.}
\label{width}
\end{figure}
%%%%%%%%%%%%

This is distinct from the effect of an extended PBH mass function, which may arise even for a delta-function power spectrum if one has critical collapse \cite{Carr:2016drx}. In this case, one needs to  consider which value of $M$ dominates the $\mu$ distortion.
This depends upon the form of the PBH mass function; however, the  limit will be important whenever this extends above $10^5M_{\odot}$, and it will be strongest if it extends up to $10^9 M_{\odot}$. The two most plausible mass functions are lognormal (which applies for a large class of inflationary scenarios) and a steeply rising power law with an exponential upper cutoff (which applies for critical collapse). In the first case the mass function only extends over a few decades of mass, so the above analysis is still a reasonable guide. In the second case, the mass function is essentially monochromatic since the number of PBHs on the power-law tail is very small. If the mass function has a power-law form, so that it is very extended, the situation is more complicated, but this is less likely to apply in any realistic scenario.  As discussed in Ref.~\cite{crtvv}, all the constraints on $f(M)$ - not just the $\mu$ constraints -  are modified anyway for an extended PBH mass function, but we do not consider this complication further here.
%One can  also consider the consequences of the PBHs having an extended mass function, a possibility which arises even if the fluctuations are monochromatic \cite{Carr:2016drx}. 
%The extended PBH mass spectrum will contribute significantly to the $\mu(M)$ distortions at large mass

\section{SUMMARY}

PBH formation is a plausible
process in the very early Universe. It is motivated observationally by the need for dark matter and by LIGO observations, and theoretically by generic inflationary fluctuation considerations.  
An inevitable consequence of PBH formation is that the dissipation of primordial fluctuations is also expected to be large. We have developed alternative formalisms for studying primordial non-Gaussianities  in these fluctuations and evaluated the associated $\mu$ limits on $f(M)$. 
%their rate of dissipation.
Our formalism is equivalent to the more conventional $f_{\mathrm{NL}}$ approach in other recent studies of PBH formation from non-Gaussian fluctuations \cite{Byrnes:2012yx}. 
We have demonstrated that  Silk damping could produce unacceptably large $\mu$ distortions of the cosmic microwave background since the fluctuations with wave numbers corresponding to the PBH masses under consideration are expected to dissipate during the redshift interval $5\times 10^4\lesssim z\lesssim2\times 10^6$. 
 We have explored current  and future constraints on these $\mu$ distortions and shown how they can be translated into upper limits on the PBH abundance over a wide mass range.  

\begin{acknowledgments}
We thank T. Harada, M. Kamionkowski, K. Kohri, V. Poulin, A. Vilenkin, J. Yokoyama and an anonymous referee for helpful input. 
T.N. and B.C. thank the Research Center for the Early Universe (RESCEU) at the University of Tokyo and the 
Institut Astrophysique de Paris for hospitality received during this work. T.N. was supported by
JSPS Postdoctoral Fellowships for Research Abroad and J. S. by European Research Council Grant
No. 267117.
 %BC thanks the Research Center for the Early Universe (RESCEU) at the University of Tokyo for hospitality received during this work. 

\end{acknowledgments}

\end{document}